\documentclass[twocolumn,pra,showpacs]{revtex4-2}
\begin{document}
\draft

\title{\bf Enhancement of weak interactions in phase transitions in condensed matter and early universe}
%, Critical Size of the New Phase Nucleus, and Implications for Baryogenesis}
\author{V.V. Flambaum}
\address{School of Physics, University of New South Wales, Sydney 2052, Australia}
\email{v.flambaum@unsw.edu.au}
\date{\today}

\begin{abstract}
Parity-violating weak interactions produce extremely small energy differences between left- and right-handed chiral systems. We show that these microscopic effects may be amplified during collective phenomena such as phase transitions. The enhancement factor is proportional to the critical number of atoms, $N_c$, in the nucleus of the new phase. Measurement of the ratio of produced left- and right-handed chiral structures may provide a way to determine $N_c$. Experiments in which spin-chiral structures are formed in ZnCr$_2$Se$_4$ during a phase transition in crossed electric and magnetic fields indicate an effective amplification corresponding to $N_c\sim 10^9-10^{10}$. We emphasize, however, that such large values cannot be inferred from homogeneous capillary nucleation alone, since the same parameters imply a very large nucleation barrier; in real crystals heterogeneous nucleation, defects, strain, and domain-growth dynamics may be essential. We also discuss, more cautiously, whether analogous collective enhancement could operate in first-order cosmological phase transitions beyond the Standard Model. The Standard-Model electroweak transition for the observed Higgs mass is a crossover and therefore has no critical bubble nucleus.

\end{abstract}

\maketitle

%--------------------------------------------------
\section{Introduction}
Parity violating (PV) weak interactions are  tiny at the energy scales of condensed matter. We demonstrate that phase transitions - specifically, nucleation at first order and large correlation volumes near criticality - provide a generic amplifier that converts microscopic PV energy differences into macroscopic chiral biases.
% The result is a simple, universal formula for the domain-population asymmetry  that collapses diverse systems onto a single scaling variable. 
 This makes parity violation experimentally accessible in chiral magnets and chiral crystallisation without external bias fields, and it yields testable predictions (e.g. handedness statistics recorded by neutron or optical probes). 
 %Beyond immediate applications, this mechanism sharpens the conceptual link between symmetry breaking in materials and in the early Universe.

As known, the Standard Model gives a baryon-to-photon ratio several  orders of magnitude smaller than observations. In this situation it is important to check that we have not overlooked any enhancement mechanism which can play a role in electroweak baryogenesis. Indeed, systems are very sensitive to weak fields during phase transitions. A very interesting experiment was performed in Ref.~\cite{Siratori1}. One may interpret its result as a \emph{nine-orders-of-magnitude} enhancement of the effects of small perturbations in phase transitions in condensed matter (see below). If a similar enhancement of CP violation existed in the electroweak phase transition (which presumably created the baryon asymmetry of the Universe), CP violation might produce a baryon-to-photon ratio compatible with observations. However, we no longer assume that the Standard-Model electroweak transition itself is a first-order nucleation process. For the observed Higgs mass it is a crossover, and the nucleation discussion can apply only to possible first-order transitions in extensions of the Standard Model.

%Even if the enhancement is not large enough in the Standard Model, it may be important for other models of baryogenesis involving a phase transition.

One may think about using enhancement of weak interaction  in phase transition to search for time reversal (T) and parity (P) violating interactions which were searched for  in the electric dipole moment experiments with neutrons, nuclei, atoms and molecules (see e.g. \cite{Khriplovich,KhriplovichLam,GFPhysRep,Mansour}). 

In nature many organic molecules prefer one chiral form over another. A striking example is that almost all DNA is right-handed, while nearly all amino acids in living organisms are left-handed (see e.g.\cite{Avalos}). One possible explanation of this phenomenon is the small energy difference between left- and right-handed molecules  (see e.g \cite{Rein,Letokhov,Bouchiat,Khriplovich}). However, it is generally believed that the parity-violating energy difference, $\Delta E_{PV}$, is too small to affect the homochirality of life.

Energy difference between molecules of different chirality is studied using spectroscopy  methods. In principle, there are other possibilities. One may study  resonance chemical  reactions and enhancement of weak interactions in collisions of cold molecules,   where chiral molecules are formed in collision of other  molecules. Weak interaction produces energy difference  $\delta E_{PV}$ in the cross section resonance position in right-hand and left-hand molecules, which results in difference in their numbers  formed \cite{FGcoldmol}. 

 In a more general case, formation of chiral molecules from non-chiral components in chemical reactions create  non-equal number of right-hand and left-hand molecules  with the relative difference which may exceed equilibrium value $\sim \exp{(- \delta E_{PV}/T)}$. This leads to the optical activity of molecular gas or solution. 
 %This would be interesting to perform a test experiment measuring difference in formation of right-hand and left-hand molecules in crossed electric and magnetic fields which also affect energy difference between the right-hand and left-hand molecules.   

A natural place to search for enhancement of such small effects is in collective phenomena such as phase transitions  - see e.g. \cite{Siratori1}. Systems are highly sensitive to weak perturbations near criticality. In the following we develop this idea within the Zeldovich nucleation model of first-order phase transitions presented e.g. in the book \cite{Landau}.

The nucleation theory may also be applied to formation of crystals.
 The  strongly enhanced effects of weak interactions may  appear in the formation of chiral crystals from concentrated solutions of non-chiral components, for example NaClO$_3$ from solutions containing ions Na$^+$ and ClO$_3^-$. Some remarkable experiments with NaClO$_3$ solutions were described in Refs.~\cite{DayPT,Viedma}. In principle, the enhancement effect in this problem may be related to the chirality of biological molecules.

%--------------------------------------------------
\section{Nucleation Model and Enhancement Factor}
In the nucleation picture, fluctuations create small droplets of a new phase. Droplets smaller than a critical size $r_{cr}$ quickly disappear, while larger ones grow to fill the system. The probability of forming a  nucleus of radius $r$ is \cite{Landau}
\begin{equation}
S \sim e^{-W_{min}/kT},
\end{equation}
where $W_{min}$ is the minimal work required to form a  nucleus,  which may be presented as a sum of the negative volume term, reflecting advantage of the new phase,  and positive surface term: $W_{min}(r)=-B \, 4 \pi r^3/3 + \alpha_t \,4 \pi  r^2$. This function has maximum at the critical size 
$r_c=2\alpha_t/B$.  This looks like $W_{min}(r)$ produces a potential barrier for formation of the nucleus bigger than the critical size, with $r>r_c$.

 Using formulas  presented in the book \cite{Landau}, it is easy to express $W_{min}(r_c)$ in terms of the chemical potentials per particle of the old and new phases, $\mu_1$ and $\mu_2$:
\begin{equation}
W_{min} = \frac{1}{2} N_c (\mu_1 - \mu_2),
\end{equation}
where $N_c$ is the number of particles in the critical nucleus.

If the new phase may exist in two nearly degenerate forms (e.g.\ left- and right-handed chirality), then weak interactions induce a small energy difference $\Delta E$ per particle. This results in slightly different $W_{min}^\pm$ for the two structures. The asymmetry in their nucleation rates can be expressed as
\begin{eqnarray}\label{P}
\nonumber
P \equiv \frac{S_+ - S_-}{S_+ + S_-} 
= \frac{ e^{-W_{min}^+/kT} - e^{-W_{min}^-/kT} }{ e^{-W_{min}^+/kT} + e^{-W_{min}^-/kT} }\\
= \tanh\!\left( -\frac{W_{min}^+ - W_{min}^-}{2kT} \right).
\end{eqnarray}
For $|W_{min}^+ - W_{min}^-| \ll kT$ the asymmetry is small and reduces to
\begin{equation}
P \approx -\frac{W_{min}^+ - W_{min}^-}{2kT}.
\end{equation}
When the phase transition proceeds from a non-chiral to a chiral phase,
\begin{eqnarray}
\label{eq:P3}
P%=-\frac{W_{min}^+ - W_{min}^-}{2kT}  
%\approx - \frac{1}{4kT}\!\left[-N_c(\mu_2^+ - \mu_2^-)+(N_c^+ - N_c^-)(\mu_1 - \mu_2)\right]\\
\sim  - \frac{N_c \Delta E}{2kT}.
\end{eqnarray}
%where $N_c = (N_c^+ + N_c^-)/2$ and $\mu = (\mu^+ + \mu^-)/2$. 
Thus, the small microscopic energy difference $\Delta E$ is enhanced by the collective factor $N_c$, the number of particles in the critical size nucleus,  which may be very large.

Here we see the potential for large enhancement: normally in statistical physics energies per degree of freedom are compared to $kT$, but a critical nucleus may contain billions of degrees of freedom, and the \emph{total} difference $W_{\min}^+ - W_{\min}^-$ is compared with $kT$.

At the same time, $N_c$ is not an independent free parameter. From Eq.~(2), $W_{min}=N_c(\mu_1-\mu_2)/2$, so a very large $N_c$ near the transition also implies a large homogeneous nucleation barrier. Efficient homogeneous nucleation on laboratory time scales usually requires $W_{min}/kT$ not to be too large. Therefore, when a very large effective $N_c$ is inferred experimentally, one must also consider heterogeneous nucleation, defects, finite-size domains, strain, or domain-growth dynamics.

Note that $N_c \propto (\mu_1 - \mu_2)^{-3}$ \cite{Landau}. The difference  in  $\mu_1 -\mu_2$ appears in metastable state of overcooled system (we assume that ordered phase appears for $T<T_c$).  Therefore, $N_C$  diverges  as $\mu_1 \to \mu_2$ for $T \to T_C$. Purely theoretically, if the temperature is  reduced very slowly, giving sufficient time for very improbable, exponentially suppressed formation of very large  nuclei of ordered phase, the enhancement may be nearly infinite, the system always comes to the lower energy chiral state.  However, this statement is challenged by the fluctuations and effects of defects.    

Another way to formulate this enhancement:  the transition temperature is different for transitions to states of different chirality.  So, in an ideal situation,  with a very slow cooling process we may achieve transition  specifically to the lower energy chiral state which has higher $T_C$. 

In higher-order phase transitions there are no metastable phases for $T<T_C$, and  transition always happens at $\mu_1 = \mu_2$.  In the first order transitions  $N_c \propto (\mu_1 - \mu_2)^{-3}$ \cite{Landau}, so one could conclude that 
the  second-order transitions can in principle provide bigger  enhancement.
% (theoretically, $N_c$ may be approaching the  number of particles in the  system  or in the  correlation volume).
However, in a genuine second-order transition or crossover there is no metastable critical droplet and hence no critical nucleus in the Zeldovich sense. Large correlation volumes near criticality may still enhance the response to small symmetry-breaking fields in condensed matter, but this is a different mechanism and should not be identified with the first-order nucleation factor $N_c$. In particular, this argument cannot by itself provide electroweak baryogenesis in the Standard Model crossover.

%{\bf In equilibrium at $T=T_c$ in first (or higher) order phase transition $\mu_1 =\mu_2$, The difference  in  $\mu_1 -\mu_2$appears in metastable state of overcooled system.}

%{\bf Formation of chiral molecules from non-chiral components create  non-equal number with relative difference $\exp{(- \delta E/T)}$ and optical activity.} 

%--------------------------------------------------
%\section{Applications in Condensed Matter}
%\subsection{Chiral Molecules}
%Alanine, the simplest amino acid, is often studied due to its biological relevance. Its $\Delta E_{PV}$ is small because it consists of light atoms, and its phase transitions often occur between two chiral states. However, $\Delta E_{PV}$ in alanine depends strongly on the torsion angle $\phi$ between the carboxylate plane and the $C_\alpha$--CO$_2$--$H_\alpha$ plane \cite{Mason,Berger,Wesendrup}. Even the sign of $\Delta E_{PV}$ changes with $\phi$. The crystalline form is believed to have $\phi \sim 62^\circ$ \cite{Lehmann}, whereas $\phi$ in aqueous solution is unknown. During a phase transition, $\mu_1^\pm$ and $\mu_2^\pm$ may differ substantially, leading to an enhancement by $N_c$.

\section{Magnetic Crystals with Spin-Screw Structures}
In  magnetic crystals the  spin-screw structures play the role of molecular chirality. The energy difference between left- and right-handed spin-screw states arises from spin-orbit coupling and external fields, see, e. g.,  book \cite{Khriplovich}.

For the spinel ZnCr$_2$Se$_4$  the phase transition is classified as a weak first order phase transition (i.e. close to the second order phase transition). Near its N\'eel temperature $T_N$=21 K, experiment in crossed electric field $E=2.5$ kV/cm and magnetic field $H=12$ kOe  measured an asymmetry $P \simeq -0.9$ \cite{Siratori1}.  The effect is produced by interactions with magnetic field  $\hat H_{H}$, electric field  $\hat H_{E}$, and spin-orbit interaction $\hat H_{\mathrm{SO}}$   \cite{Siratori1,Khriplovich}. A rough estimate yields 
\begin{eqnarray}\label{EHE}
\Delta E \sim  \frac{\langle n|\hat H_{\mathrm{SO}}|m\rangle \langle m|\hat H_{E}|k\rangle \langle k|\hat H_{H}|n\rangle}{(E_n^{(0)}-E_m^{(0)})(E_n^{(0)}-E_k^{(0)})}\,\, \\
\sim \eta_{EH} \frac{\, Z^2 \alpha^2 (e^2/a_B)\cdot e a_B E \cdot \mu_B H}{(e^2/a_B)^2}
\sim \eta_{EH} \,\, 10^{-12} \mathrm{eV}\,\, \\
  \approx  \eta_{EH} \,\,10^{-8} \mathrm{K} \,\,  .
% -1.5  10^{-13} \mathrm{a.u.} - 4 10^{-12} \mathrm{eV}.
\end{eqnarray}
Geometric suppression factor $\eta_{EH} <1 $  may reduce this estimate. Using Eq.  (\ref{eq:P3}) we obtain
%indicating that the left-handed helix has the lower energy. Choosing $T > T_N$ (but $T \simeq T_N$) ensures $\mu_1 < \mu_2$ and yields the bound
\begin{equation}
N_c \gtrsim   10^{9}.
%N_c \gtrsim  1.5 \times 10^{9}.
\end{equation}
%If so, the weak-interaction contribution alone would be enhanced to a measurable level:
%\begin{equation}
%P \sim \frac{\Delta E_{PV}}{4kT_N}\times 1.5\times 10^9.
%\end{equation}
%Although optimistic (geometric suppression factors may be important),

 Since the observed asymmetry is large, the exact expression (\ref{P}) should be used rather than the linear approximation. If one nevertheless parametrizes the observation by an effective critical number, then
\begin{equation}
N_c^{\rm eff}=
\frac{2kT_N}{|\Delta E_{EH}|}\tanh^{-1}|P|
\simeq 5 \times10^9
\left(\frac{10^{-12}\,{\rm eV}}{|\Delta E_{EH}|}\right).
\end{equation}
Thus the data correspond phenomenologically to $N_c^{\rm eff}\sim 10^9-10^{10}$ for $|\Delta E_{EH}|\sim 10^{-12}$ eV. This should not be interpreted as a proof that homogeneous capillary nucleation occurs with this value of $N_c$. As shown in the Appendix, homogeneous nucleation with $N_c\sim 10^9$ would have an enormous barrier. The experiment therefore motivates the enhancement mechanism, but a quantitative microscopic explanation of the observed domain selection requires a model of heterogeneous nucleation, defects, strain, or non-equilibrium domain growth.

 Note that heterogeneous nucleation typically rescales
 $W_c$ and $N_c$ by the same shape factor,
 so it does not obviously lift the tension between
 a large $N_c$ and efficient nucleation.

%This illustrates the potential scale of the enhancement.

\section{Effect of parity violating weak interaction}

The estimate of the collective enhancement factor \(N_c\) from the
Zeldovich nucleation theory \cite{Landau} is not unique, since several
material parameters are not known accurately. In particular, the result
depends strongly on the interfacial energy and on the actual cooling
protocol, which determines the undercooling temperature. 
 As discussed in
the Appendix, a capillary estimate gives $N_c\sim10^9$ only for very small
undercooling.  
%one may nevertheless conclude that critical nuclei
%containing \(10^9\) spins are plausible in ZnCr$_2$Se$_4$ if the
%transition is weakly first order and occurs sufficiently close to  $T_N$. 
However, the corresponding homogeneous nucleation
barrier is then extremely large. Thus this value should be interpreted as
an effective collective scale suggested by the observed chirality
selection, not as a self-contained homogeneous nucleation calculation.

%Our attempt to do numerical  estimate of the collective enhancement factor $N_c$ using Zeldovich theory \cite{Landau}  has not given a definite answer since several unknown parameters are involved (see Appendix). Moreover, specific value of $N_c$ depends on the cooling protocol which determines overcooling temperature.  We only can conclude that 
% critical nuclei containing $10^9$ spins are quite plausible in ZnCr$_2$Se$_4$ if the transition is weakly first order and occurs sufficiently close to $T_N$. 

 We may estimate the ratio of the effect produced by the parity violating  weak interaction (in the absence of external fields) to the  observed effect   in the crossed electric and magnetic fields.
 PV electron-nuclear interaction can be written as a sum of the nuclear-spin-independent (NSI) and nuclear-spin-dependent (NSD) parts:
\begin{equation} \label{HPV}
  H_{PV}  = \frac{G_F}{\sqrt{2}} \left( - \frac{Q_W}{2} \gamma_5 + 
 \frac{\kappa}{I}   {\bf \alpha}\cdot{\bf I}\right) \rho(r),
\end{equation} 
where $G_F$ is the Fermi weak interaction constant, $Q_W$ is the nuclear weak charge, $\alpha $ and
 $\gamma_5$ are the Dirac matrices, $I$ is the nuclear spin, and $\rho(r)$ is the nuclear density normalised to 1.
The effective nuclear weak  charge, including radiative corrections, is equal to \cite{FlambaumSamsonov} 
\begin{equation}
    Q_W = -0.98897\,N + 0.070605 \,Z. 
\end{equation}
Here $N$ is the number of neutrons, and $Z$ is the number of protons.
The dimensionless constant $\kappa \sim 1$ is dominated by the interaction of electrons with the nuclear anapole moment  and determines the strength of the NSD PV interaction \cite{FlambaumKhriplovich} . 

The energy difference $\Delta E_{PV}$  between opposite
enantiomers  originates from the PV interaction (\ref{HPV}) between electrons and nuclei. The weak interaction mixes $s_{1/2}$ and $p_{1/2}$ orbitals, creating a spin-helical structure within atoms  \cite{Bouchiat,Khriplovich}). The spin-orbit interaction then distinguishes between coordinate and spin structures aligned in opposite directions.
Estimate gives \cite{ReinHegstromSandars1979,HegstromReinSandars1980}
\begin{equation}\label{EPV}
\Delta E_{PV} \sim  10^{-19} Z^{5}\eta _{PV} \quad \mathrm{eV},
\end{equation}
where $Z$ is the nuclear charge of the heaviest atom in the molecule and $\eta_{PV}$ is a molecular asymmetry factor. The steep $Z^5$ dependence results from weak ($\propto Z^3$) and spin-orbit ($\propto Z^2$) scaling, but $\eta$ is typically small.  Comparison with accurate calculations \cite{Mason,Bakasov,Laerdahl,Berger,Wesendrup,QuackStohner2000PRL,Viglione2004JCP,RauhutSchwerdtfeger2021PRA,Sunaga2025arXiv,Eduardus2023CC,LaerdahlSchwerdtfegerQuiney2000PRL,BastSchwerdtfeger2003PRL,RauhutBaroneSchwerdtfeger2006JCP} indicate $\eta_{PV} \sim 10^{-5}$.  In molecules with two heavy atoms, $\eta_{PV}$ can be few orders of magnitude larger than in molecules with one heavy atom \cite{ReinHegstromSandars1979,HegstromReinSandars1980}. 
%For light molecules such as alanine $\Delta E_{PV}\sim 10^{-21\pm 3}$ a.u.~\cite{Mason,Bakasov}, while
In molecules containing heavy atoms  $\Delta E_{PV}$ may exceed $10^{-11}$ eV, see e.g.  \cite{Mason,Bakasov,Laerdahl,Berger,Wesendrup,QuackStohner2000PRL,Viglione2004JCP,RauhutSchwerdtfeger2021PRA,Sunaga2025arXiv,Eduardus2023CC,LaerdahlSchwerdtfegerQuiney2000PRL,BastSchwerdtfeger2003PRL,RauhutBaroneSchwerdtfeger2006JCP}.

Now we can estimate the ratio of the energy difference produced by the parity violating interaction (\ref{EPV}) to  that produced by crossed electric and magnetic field  (\ref{EHE}): 
\begin{equation}
       \frac{ \Delta E_{\rm PV}}{\Delta E_{EH}} \sim    10^{-4} Z^3 \frac{\eta_{PV}}{\eta_{EH}},
        \label{eq:ratioestimate}
\end{equation}
  For ZnCr$_2$Se$_4$, the largest nuclear charge is $Z=34$, and therefore
\begin{equation}
       \frac{ \Delta E_{\rm PV}}{\Delta E_{EH}} \sim     \frac{\eta_{PV}}{\eta_{EH}},
        \label{eq:ratiospinel}
\end{equation}
Thus, if the structural factors for the PV and crossed-field mechanisms
are not too different, the PV interaction in systems with a heavy atom may produce a bias comparable
to, or even larger than, the crossed-field effect observed in
Ref.~\cite{Siratori1}. This estimate is only qualitative: in a magnetic
crystal the relevant ``chirality'' is the spin-screw chirality rather
than ordinary molecular chirality, and the factors \(\eta_{PV}\) and
\(\eta_{EH}\) require a microscopic calculation. Nevertheless,
Eq.~(\ref{eq:ratiospinel}) suggests that parity-violating weak
interactions in heavy magnetic crystals could be observable through the
same phase-transition enhancement mechanism.

%More accurate treatment of this problem requires numerical calculations which are beyond the scope of the present paper.

%In holmium, Fedorov {\it et al.}~\cite{Fedrov} observed $P < 6\times 10^{-5}$, much smaller than the spinel case. Using Khriplovich’s estimate $\Delta E \simeq 1.2\times 10^{-14}$ a.u.\ gives $N_c < 8.3\times 10^6$, consistent with limited enhancement.

%--------------------------------------------------
\section{Possible implications for baryogenesis}
Several studies have concluded that the Standard Model predicts a baryon-to-photon ratio several orders of magnitude smaller than observed in electroweak-baryogenesis estimates (see, e.g,  \cite{Trodden,Gavela,Huet,Farrar,FlambaumShuryak}). It is therefore useful to ask whether collective effects during phase transitions could amplify CP-violating biases.

 The Standard-Model electroweak transition for the observed Higgs mass is a crossover, not a first-order phase transition. Therefore there is no critical bubble nucleus in the Standard Model electroweak transition. The nucleation argument discussed in this paper can apply to cosmology only in models beyond the Standard Model in which the electroweak or another cosmological transition is first order.

In ordinary electroweak baryogenesis, bubble walls separate the high-temperature symmetric phase from the low-temperature broken phase. CP-violating interactions modify the transport of particles and antiparticles in front of a moving wall, while sphaleron processes in the symmetric phase convert the resulting chiral asymmetry into baryon number (see, e.g., Refs.~\cite{Trodden,Farrar,Gavela,Huet,FlambaumShuryak}). 
%{\bf The relevant bubbles are bubbles of the broken Higgs phase; they should not be described as separate ``matter'' and ``antimatter'' bubbles.}

For a first-order transition, the temperature at which bubbles actually nucleate is not exactly the critical temperature.   In cosmological applications it is determined by the condition that the nucleation rate becomes comparable with the Hubble expansion rate.
%conventionally expressed as $\Gamma/H^4\sim1$ or $S_3(T_n)/T_n\simeq 140$. 
The transition then completes by bubble growth, collision, and percolation, rather than by a single bubble filling the Universe. The amount of supercooling, the bubble-wall velocity, the surface tension, and the latent heat are model dependent.

The condensed-matter mechanism discussed above therefore should be viewed only as an analogy and motivation for further study.  It does not constitute a quantitative theory of baryogenesis. A quantitative cosmological calculation would require a specific first-order model beyond the Standard Model, together with the finite-temperature effective potential, bubble action, wall profile, hydrodynamics, and CP-violating transport. Nevertheless, the analogy suggests a question worth investigating: whether CP-violating effects that modify bubble-wall transport, pressure balance, or the critical action could be amplified by collective properties of a first-order transition.

The nucleation framework and its $N_c$-based enhancement is not directly  applicable to  a crossover. Note however, that  even in a crossover, droplet-like regions ("bubbles") of the emergent  phase appear  within the symmetric phase; see, e.g., \cite{FlambaumShuryak} and references therein. Therefore, a possibility of a collective enhancement in the crossover case may also deserve  investigation.

%--------------------------------------------------
\section{Conclusion}
%Weak-interaction--induced parity-violating energy differences are generally tiny, but during phase transitions they may be amplified by large collective factors $N_c$.
%This mechanism could contribute to fundamental asymmetries such as molecular homochirality and baryogenesis. 

 We have discussed a simple mechanism by which very small energy differences may be amplified during phase transitions. In the Zeldovich nucleation picture the probability of forming a critical nucleus depends exponentially on the minimum work \(W_{\min}\). Therefore, if two possible structures differ in energy by a small amount \(\Delta E\) per particle, the asymmetry in their formation probabilities contains the collective factor \(N_c\), the number of particles in the critical nucleus:
%We have analyzed a simple mechanism by which energy differences are amplified during phase transitions. In the Zeldovich nucleation framework the macroscopic bias
\[
P \sim -\frac{N_c\,\Delta E}{2kT}.
\]
This formula links a microscopic splitting $\Delta E$ to a collective factor $N_c$ - the number of particles in the critical nucleus of ordered phase. Enhancement factor $N_c$ grows rapidly as the temperature of overcooled material approaches critical transition temperature, $T \to T_c$,   and, in principle, could become extremely large.
% in weakly first-order or effectively second-order cases. 
 However,  $N_c$ is tied to the nucleation barrier and cannot be chosen independently. Large $N_c$ values are possible near a weakly first-order transition, but homogeneous nucleation with such large nuclei may be exponentially suppressed. Therefore, the large effective enhancement suggested by ZnCr$_2$Se$_4$ should be regarded as phenomenological unless heterogeneous nucleation, defects, strain, or domain-growth dynamics are included.

This mechanism suggests experimental searches for parity-violating effects based not only on high-resolution spectroscopy, but also on measurements of chiral domain statistics or optical activity generated during crystallisation. A similar mechanism may enhance the difference in the formation of chiral crystals from concentrated solutions of non-chiral components. Measurement of the ratio of produced left- and right-handed chiral structures, combined with calculations of the corresponding microscopic energy differences, may provide a way to determine an effective collective enhancement factor.

 The possible cosmological implications have also been mentioned. The Standard-Model electroweak transition is a crossover and has no critical bubble nucleus. Hence the first-order nucleation mechanism considered here can be relevant to cosmology only in appropriate extensions of the Standard Model. A quantitative baryogenesis calculation would require a specific finite-temperature model, including the bubble action, wall profile, hydrodynamics, sphaleron processes, and CP-violating transport. 
In a finite condensed-matter sample a supercritical nucleus grows until it collides with other domains or reaches the sample boundary; in a cosmological first-order transition the transition proceeds by nucleation, growth, collision, and percolation of many bubbles.

Thus, the main result of the present work is a general amplification mechanism and its order-of-magnitude consequences, rather than a complete microscopic calculation for a particular material or a quantitative theory of baryogenesis. The estimates presented here motivate detailed numerical calculations and new measurements in systems where small parity-violating energy differences may be converted into observable macroscopic asymmetries.

\vspace{2mm}
\textit{Acknowledgements.}--- 
I am grateful to Elizabeth Angstmann and Jacinda Ginges  for their  contributions during the early  stages of this work, and to Bodi Shaibat for his help in estimate of the collective enhancement factor $N_c$ in real materials.
 The work was supported by the Australian Research Council Grant No.\ DP230101058.

 %--------------------------------------------------
\appendix
\section{Estimate of the critical nucleus size in ZnCr$_2$Se$_4$}

In this Appendix we give an order-of-magnitude  estimate of the critical size of the nucleus of the helical phase in ZnCr$_2$Se$_4$. The estimate is intended only to show that a very large critical number of spins, $N_c$, is plausible near a weak first-order phase transition.

The free energy of a spherical nucleus of radius $r$ is
\begin{equation}
        W(r)=4\pi r^2\alpha_t-\frac{4\pi}{3}r^3 B ,
        \label{eq:capillary_W}
\end{equation}
where $\alpha_t$ is the interfacial energy between the paramagnetic and helical phases, and $B$ is the bulk free-energy gain per unit volume. The critical radius is determined by $dW/dr=0$, giving
\begin{equation}
        r_c=\frac{2\alpha_t}{B}.
        \label{eq:critical_radius_general}
\end{equation}
Close to the transition temperature, the bulk driving force may be written as
\begin{equation}
        B\simeq \Delta s_v\,\delta T ,
        \label{eq:bulk_driving_force}
\end{equation}
where $\Delta s_v$ is the entropy change per unit volume and
\[
        \delta T=T_N-T
\]
is the undercooling below the transition temperature. Therefore,
\begin{equation}
        r_c=\frac{2\alpha_t}{\Delta s_v\,\delta T}.
        \label{eq:critical_radius}
\end{equation}
The corresponding number of Cr spins in the critical nucleus is
\begin{equation}
        N_c=
        \frac{4\pi}{3} n_{\rm Cr} r_c^3
        =
        \frac{4\pi}{3} n_{\rm Cr}
        \left(
        \frac{2\alpha_t}{\Delta s_v\,\delta T}
        \right)^3 ,
        \label{eq:Nc_capillary}
\end{equation}
where $n_{\rm Cr}$ is the density of Cr spins.

The spinel unit cell contains eight formula units and therefore 16 Cr ions. Taking the lattice constant
\[
        a\simeq 10.5\ {\rm \AA},
\]
one obtains
\begin{equation}
        n_{\rm Cr}=\frac{16}{a^3}
        \simeq 1.38\times 10^{28}\ {\rm m^{-3}} .
        \label{eq:nCr}
\end{equation}
The molar volume is
\begin{equation}
        V_m=\frac{N_Aa^3}{8}
        \simeq 8.7\times 10^{-5}\ {\rm m^3\,mol^{-1}} .
        \label{eq:molar_volume}
\end{equation}
Using the entropy change at the transition,
\begin{equation}
        \Delta S\simeq 1.8-3.1\ {\rm J\,mol^{-1}K^{-1}},
\end{equation}
gives
\begin{equation}
        \Delta s_v=\frac{\Delta S}{V_m}
        \simeq (2.1-3.6)\times 10^4\ 
        {\rm J\,m^{-3}K^{-1}} .
        \label{eq:entropy_density}
\end{equation}

The least certain parameter is the interfacial energy $\alpha_t$. Since no direct measurement of the interface energy between the paramagnetic and helical phases appears to be available, we estimate its scale from the micromagnetic expression
\begin{equation}
        \alpha_t\sim 4\sqrt{A_{\rm ex}K_u},
        \label{eq:alpha_estimate}
\end{equation}
where $A_{\rm ex}$ is the exchange stiffness and $K_u$ is the anisotropy energy density. Taking
\begin{equation}
        A_{\rm ex}\sim (1-3)\times 10^{-12}\ {\rm J/m},
        \qquad
        K_u\sim (3-7)\times 10^3\ {\rm J/m^3},
\end{equation}
one obtains the order-of-magnitude range
\begin{equation}
        \alpha_t\sim (2-6)\times 10^{-4}\ {\rm J\,m^{-2}} .
        \label{eq:alpha_range}
\end{equation}

Substitution of Eqs.~(\ref{eq:nCr}), (\ref{eq:entropy_density}), and
(\ref{eq:alpha_range}) into Eq.~(\ref{eq:Nc_capillary}) gives
\begin{equation}
        N_c \simeq
        (0.08-11)\times 10^6
        \left(
        \frac{1\ {\rm K}}{\delta T}
        \right)^3 .
        \label{eq:Nc_range}
\end{equation}
A representative central value is
\begin{equation}
        N_c \simeq
        1.4\times 10^6
        \left(
        \frac{1\ {\rm K}}{\delta T}
        \right)^3 .
        \label{eq:Nc_central}
\end{equation}

Thus the critical nucleus size is extremely sensitive to the undercooling:
\[
        N_c\propto \delta T^{-3}.
\]
For example, if the transition occurs with an undercooling of order
\[
        \delta T\sim 0.1\ {\rm K},
\]
Eq.~(\ref{eq:Nc_central}) gives
\[
        N_c\sim 10^9 .
\]
The corresponding critical radius is
\begin{equation}
        r_c=
        \left(
        \frac{3N_c}{4\pi n_{\rm Cr}}
        \right)^{1/3}
        \sim 3\times 10^{-7}\ {\rm m},
\end{equation}
i.e. a few tenths of a micron.

 We now add an important qualification. The same capillary parameters also determine the homogeneous nucleation barrier,
\begin{equation}
        W_c=\frac{16\pi \alpha_t^3}{3B^2}
        =\frac{16\pi \alpha_t^3}{3(\Delta s_v\delta T)^2} .
        \label{eq:Wc_barrier}
\end{equation}
For the central values used above and $\delta T\simeq0.1$ K one finds
\begin{equation}
        \frac{W_c}{kT}\sim 5\times10^5,
\end{equation}
so homogeneous thermal nucleation would be exponentially suppressed. Conversely, requiring an efficient homogeneous nucleation rate, roughly $W_c/kT\sim40$--$100$ on laboratory time scales, 
%gives  a much larger undercooling, $\delta T\sim7$--$11$ K for the central parameters. Eq.~(\ref{eq:Nc_central}) 
would then give only
\begin{equation}
         N_c\sim10^3-10^4,
\end{equation}
 and the asymmetry produced by a microscopic energy difference $\Delta E\sim10^{-12}$ eV at $T\simeq21$ K would be only
\begin{equation}
         |P|\sim \frac{N_c\Delta E}{2kT}\lesssim10^{-6}-10^{-5},
\end{equation}
far below the observed $|P|\simeq0.9$. Therefore the simple homogeneous capillary estimate cannot by itself explain the ZnCr$_2$Se$_4$ experiment. The large value $N_c\sim10^9$ should instead be regarded as an effective collective scale inferred phenomenologically from the observed chirality selection. A quantitative explanation requires additional physics, such as heterogeneous nucleation on defects or surfaces, strain, domain-wall motion, sample history, or other non-equilibrium mechanisms.

This estimate is necessarily approximate because the interfacial energy $\alpha_t$ and the actual undercooling in the experiment are not known accurately.  Its main purpose is to display the tension between large critical nuclei and efficient homogeneous nucleation, and to identify the additional material physics that must be included in a realistic calculation.

%--------------------------------------------------

\end{document}